# DC and AC Characterization of Pancake Coils Made from Roebel-Assembled Coated Conductor Cable

Anna Kario, Michal Vojenčiak, Francesco Grilli, Andrea Kling, Alexandra Jung, Jörg Brand, Andrej Kudymow, Johann Willms, Uwe Walschburger, Victor Zermeno and Wilfried Goldacker

*Abstract*—**Roebel cables made of HTS coated conductors can carry high currents with a compact design and reduced AC losses. They are therefore good candidates for manufacturing coils for HTS applications such as motors and generators.**
**In this paper we present the experimental DC and AC characterization of several coils assembled from a 5 meter long Roebel cable built at KIT, which differ in the number of turns and turn-to-turn spacing. Our experiments, supported by finite-element method (FEM) calculations, show that a more tightly wound Roebel coil, despite having a lower critical (and therefore operating) current, can produce a higher magnetic field than a loosely wound one. For a given magnetic field produced at the coil's center, all the coils have similar AC losses, with the exception of the most loosely wound one, which has much higher losses due to the relatively large current needed to produce the desired field.**
**The experiments presented in this paper are carried out on the geometry of pancake coils made of Roebel cables, but they are exemplary of a more general strategy that, coupling experiments and numerical simulations, can be used to optimize the coil design with respect to different parameters, such as tape quantity, size, or AC loss – the relative importance of which is dictated by the specific application.**

## I. INTRODUCTION

SECOND-GENERATION coated conductor tapes are most promising high-temperature superconductors for applications because of their current-carrying potential, mechanical flexibility, and in-field behavior. The development of coated conductors has been constantly improving in terms of pinning quality as well as of long length production[1]. Following this development, in our group work with Roebel Assembled Coated Conductor (RACC) cable has been steadily progressing [2-4]. Besides improving the single RACC cable current performance, we are also focusing on other equally important aspects: extending the cable length, reducing the AC losses, and assembling coils for possible applications. Pancake coils from coated conductors have been

A. Kario, F. Grilli, A. Kling, A. Jung, J. Brand, A. Kudymow, J. Willms, U. Walshburger, V. Zermeno and W. Goldackerare with the Karlsruhe Institute of Technology (anna.kario@kit.edu). M. Vojenčiak is now with the Institute of Electrical Engineering Slovak Academy of Science. Funding from the Helmholtz Association (FG, VMRŽ and MV, Grant VH-NG-617) and EFDA (MV, Grant WP11-FRF-KIT) is gratefully acknowledged.

studied by several groups [5-9], but until now only one pancake coil with Roebel cable was successfully made by Jiang *et al*.[10]. That coil was made from a 5 m-long Roebel cable with 9 strands, each 2 mm wide, with a 5.44 mm distance between the turns. AC loss measurement in the pancake coil wound with the Roebel cable showed hysteretic features and the AC loss was approximately 60% higher than that of a straight Roebel cable.

For the design of future Roebel cable coils it is important to understand what the most important parameters for obtaining a given magnetic field are and to have techniques and tools able to predict the coil's performance[11]. In the present work we modified the distance between the turns of the coils and investigated its influence on the performance of the coils, in terms of critical current, magnetic field produced at the center, and AC losses. Experiments are supported by FEM calculations able to compute the magnetic field distribution in detail and to estimate the coil's critical current.

## II. COIL PREPARATION AND EXPERIMENTAL METHODS

A 5 m long RACC cable was produced to assemble 5 pancake coils. Coated conductor tape from Superpower with a width of 12.01 mm, an overall thickness of 0.098 mm including a Cu stabilization thickness of 40 μm, and an average self-field critical current of 348 A (at 77 K) was used. A 50 m long piece of tape was cut into a 5.5 mm wide meander-shaped strand using our customized reel-to-reel punching tool [2]. The strand material was then cut in ten 5 m-long pieces, which were manually assembled to form a 10-strand RACC cable. A fiberglass epoxy laminate (G10) sample holder was built and used to change the distance between the turns using Styrofoam spacers (Fig.1).

Rounded copper contacts were soldered at the ends of the RACC cable using Sn60Pb40 solder at 250 °C (Fig.1). The contacts were specially designed to prevent damages to the cable due to different winding events and to be indifferently used with the various coils. Five coils with different space between the windings were prepared from the same RACC cable and characterized. In order to prevent movements due to the Lorentz force, during the measurements the coil was secured between two G10 plates. For the loosely wound coils (turn-to-turn spacing of 20, 10, 4 mm) Styrofoam material was used as spacer. For the tightly wound coils (turn-to-turn spacing of 1.4 and 0.1 mm) the turns were separated by means of a paper spacer and Kapton tape, respectively. For all coils the inner diameter was kept constant and equal to 9.2 cm; the number of turns and the outer diameter varied correspondingly. The main properties of the coils are summarized in Table I.

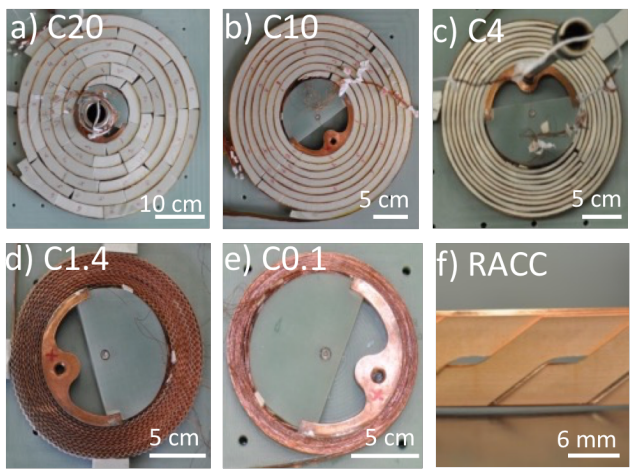

Fig. 1. Assembled coils with 20 (a), 10 (b), 4 (c), 1.4 (d) and 0.1 mm (e) separation between the turns. The massive copper blocks are the current leads. Top view of the Roebel cable (f) is also shown.

TABLE I
COIL DIMENSIONS AND SPACING MATERIAL.

| Coil | C20 | C10 | C4 | C1.4 | C0.1 |
|---|---|---|---|---|---|
| Spacing (mm) | 20 | 10 | 4 | 1.4 | 0.1 |
| Spacing material | Styrofoam | Styrofoam | Styrofoam | paper | Kapton |
| No. of turns | 6 | 9 | 9 | 11 | 13 |
| Outer diameter (mm) | 350 | 294 | 186 | 148 | 123 |

All coils were wound with the superconducting layer in the Roebel cable facing away from the coil center, so that voltage taps can be easily soldered on the superconductor's side of the tape. Cable bending experiments with the superconducting layer inwards and outwards with diameter of 92 mm (diameter of the innermost turn of the coils) were performed and no degradation of the critical current was observed. The other turns of the coils have a diameter larger than 92 mm (and a correspondingly lower tension), so one can assume that no degradation of $I_c$ occurs.

Each coil sample was characterized in DC and AC as follows. In order to measure the critical current, voltage taps were placed on each tape outside the current contacts, on the superconductor side of a single meander tape. For comparison, voltage contacts were also placed inside the coil on two tapes, which resulted in the same $I_c$, in agreement with the findings of experiments conducted on single tapes[12]. The transport AC losses were measured with the standard lock-in technique. The voltage signal was taken from the taps on one strand, as signals on all 10 strands were found to be comparable. The AC current was generated by a power amplifier and by a copper transformer cooled in liquid nitrogen.

### III. DC CHARACTERIZATION

Figure 2 shows the critical current (1 μV/cm criterion) as a function of the distance between the turns in the different coils. Reducing the distance between the turns decreases the coil's critical current. This is the result of the generated magnetic self-field, which increases when the coil is more tightly packed. The flux-creep exponent $n$ was extracted from the current-voltage relation adopting a power-law behavior: $V(I)=V_c(I/I_c)^n$. We also observed a decrease (from 24 to 18) of the power exponent with reduced turn-to-turn spacing, with the exception of coil C4. For comparison, the critical current and the power exponent of the straight RACC cable are 1002 A and 23, respectively.

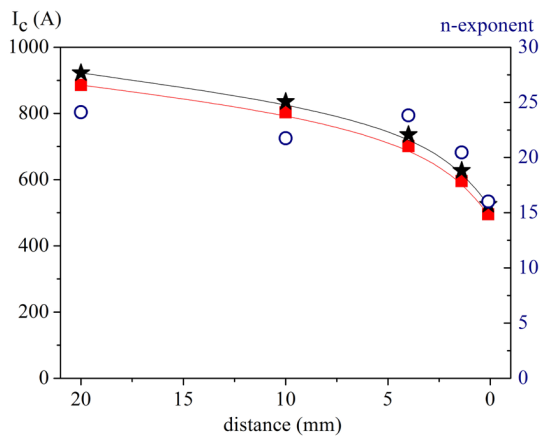

Fig. 2. Critical current (black stars are simulations, red square experimental results, lines are guide for eyes) and n factor (empty circles) as a function of the distance between turns in the coil.

The critical current of each coil was also estimated by means of FEM simulations, with a new modeling approach recently developed for this purpose [13]. The model is based on the asymptotic limit of Faraday's equation when the time $t$ approaches infinity, it includes the angular dependence of the superconductor's critical current density, and is able to calculate the voltage drop in each conductor and turn composing the coil. The voltage to determine the critical current of the coil was calculated by averaging the voltage drops of the strands in each turn and by summing these averages, taking into account their different length. The estimated voltage drop per meter in each turn is shown in figure 3. For reference, the calculated current in each coil turn with respect to local $I_c$ is also shown. Being exposed to a higher magnetic field, the innermost turns in every coil shows the largest voltage drop, hence largely contributing to the current limitation. This can be seen directly from the right axis

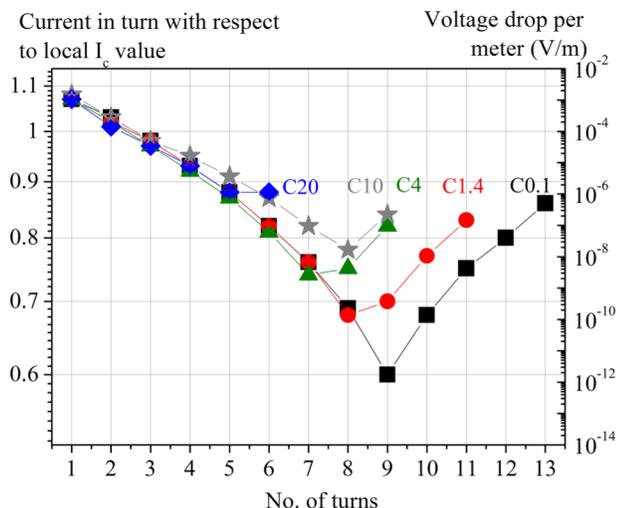

Fig. 3. Calculated current in each coil turn with respect to local $I_c$ (left axis). Voltage drop per meter in each turn (right axis).

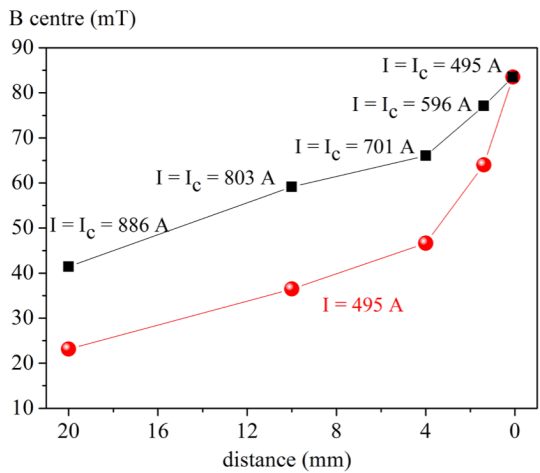

Fig. 4. Magnetic field in the center of each coil as a function of the distance between turns and critical current (black), and normalized to the critical current of C0.1 (red).

that shows a net current in the innermost turns about 7% above the local $I_c$ value.

The critical currents calculated with the FEM model are in excellent agreement (4-6 %) with the experimental values, as shown in Fig. 2, which is completely satisfactory, given the large number of effects that can make real coils different from the modeled ones (e.g. lateral and longitudinal tape uniformities, errors in the determination of angular dependence, misalignment in the Roebel structure).

If, in a given application, the coil needs to be designed to produce a field as large as possible at its center, there are two competing effects that need to be taken into account. One effect is that, in order to produce a high field at its center, the coil needs to be wound as compactly as possible. The second effect is that the critical current in a tight coil is lower than in a loosely wound coil, because of the higher produced self-field. A reduced critical current means a reduced maximum current that can be injected, and ultimately a reduced maximum generated field. FEM analysis allows evaluating the magnetic field produced by each coil, and therefore to determine, which of the two competing effects is dominant.

Figure 4 shows the calculated magnetic field produced by each coil at its center with a current equal to the critical one (black points). Coil C0.1, in spite of having a critical current of only 495 A, creates a magnetic flux two times higher (84 vs. 41 mT) than that produced by coil C20 with its critical current of 886 A. With a transport current of 495 A, coil C20 produces a field of only 23 mT – see red circle data points.

For the coils considered here, winding the cable with a small turn-to-turn spacing is advantageous not only in terms of the maximum field produced at the coil's center, but also in terms of field produced at a given applied current. In the case of other coils with different geometries or cable strands with a different angular $J_c(B)$ dependence the conclusion can be unlike. However, these results show the general effectiveness of our method for optimizing the coil design with respect to the field produced at its center.

## IV. AC Characterization

Figure 5 shows the dependence of AC losses at 36 Hz on the transport current amplitude for coils with different separation between the turns. The loss of the straight cable is similar to the loss predicted by Norris' model for a conductor with elliptical cross-section. Similar behavior has also been reported in earlier works [3][10]. As discussed above, winding the cable into a coil changes the magnetic field experienced by the superconductor: the tighter the coil is wound the higher this field is (see also for example figures 4-5 in[14]). This has an important influence on the AC loss as well, which increases with respect to that of the straight cable: in the case of the loosely wound coil (spacer 20 mm) only slightly, while in case of densely packed coil by more than 50 times. As discussed in [15] and confirmed by measurements in a dedicated study conducted on single tapes [12], the measured loss value is also influenced by the dissipation in the copper contact.

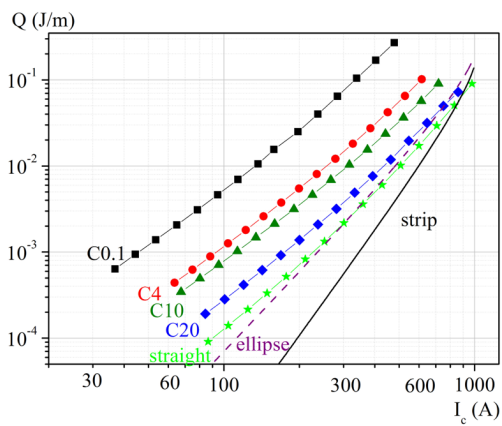

Fig. 5. Measured AC loss (36 Hz) as a function of the transport current amplitude for the different coils and for the straight Roebel cable. Norris' analytical prediction for straight conductors with $I_c$=1002 A are also shown.

As seen from the results reported above, the coil with the thinnest spacer has the highest loss. However, this coil produces also the highest magnetic field at a fixed current. In coil applications the maximum reachable magnetic flux density is usually the parameter around which the coil is tailored. Therefore one can think of representing the AC losses as a function of this parameter, instead of the transport current. For this purpose, we used the calculated flux density in the center of the coil as independent parameter – see Fig. 4. In this representation the loss curves are mostly overlapping (Fig.6.). We found a small, non-monotonic dependence of the loss on the distance between turns at fixed magnetic flux density. This dependence for a field of 35 mT is shown in figure 7.

Following this approach one can optimize the spacer thickness to reduce the losses at a given magnetic field. C4 has the lowest losses at a given field, but only by a small margin (10 %) with respect to C10 and C0.1. On the contrary coil C20 has much higher losses (94 % higher) at the same given field. By increasing the spacing, the coil is composed of less turns: as a consequence, in order to produce a given field, more current needs to be injected, which increases the losses. This is caused by the fact that with a large separation, the coil tends to increasingly behave like an isolated Roebel cable.

## V. Conclusions

We manufactured a 5 m long RACC cable with high current capacity (1002 A) and used it to assemble pancake coils with different distances between the turns: 20, 10, 4, 1.4 and 0.1 mm. We investigated experimentally the influence of the turn separation on the critical current and on the AC losses of those coils. The pancake coil geometry using Roebel cables is interesting for applications such as motors and generators where space and AC losses are important issues. We also used FEM analysis to calculate critical currents of the coils and the magnetic field produced at their center. The results of $I_c$ calculation are in excellent agreement with the experimental characterization, showing an average difference of 5% for the value of the coil's critical current.

The results allowed us to evaluate the trade-off between two competing effects: winding tighter coils enables reaching higher magnetic fields, but it also reduces the critical current of the coil and increases its AC losses. For the coils considered in this paper, in the case of DC applications the first effect dominates: the most tightly wound coil (0.1 mm turn separation) allows reaching higher fields despite the lower current that can be injected into it. However, for AC applications, the coil with the lowest losses at a given produced field is the one with 4 mm turn separation, but only with 9% and 10% differences with respect to C10 and C0.1 respectively.

This paper presents results for a specific case, but can be seen as an example of a general strategy for optimizing the coil design. That kind of numerical analysis is not only limited to coils that might be simplified to 2-D, but also other coil geometry like racetrack [16]. Our results, based on coupling experimental characterization and FEM analysis, are exemplary of an effective strategy to find the design of a coil that can generate, for example, a given magnetic field with the lowest AC losses. Other parameters important in applications, such as the size of the coil or the quantity of superconducting tape used, can be entered in the quest for optimization.

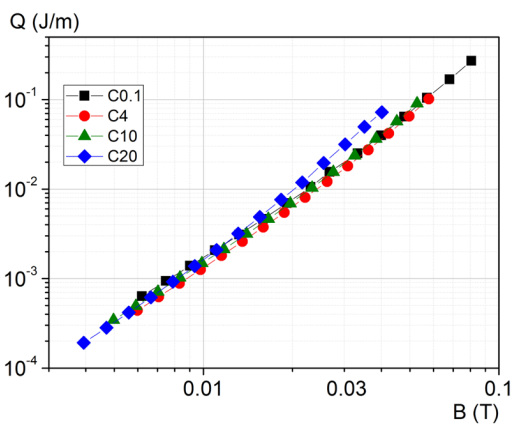

Fig. 6. AC losses as a function of the magnetic field at the center of coils with different distance between turns.

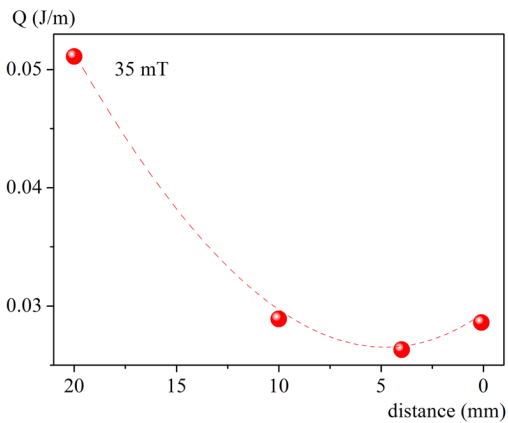

Fig. 7. AC losses vs. distance between turns of each coil for a central magnetic field of 35 mT at 36 Hz.